\documentclass[prb,aps,twocolumn,psfig,showpacs]{revtex4}
\usepackage{epsfig}
\usepackage{graphicx}
\usepackage{amsmath}

\begin{document}

\title{Structural and dynamical properties of  
liquid Si. An orbital-free molecular dynamics study}

\author{A. Delisle$^{1}$, D.J. Gonz\'alez$^{2}$ and M.J. Stott$^{1}$}

\affiliation{$^{1}$ Department of Physics, Queen's University, 
Kingston, Ontario, CANADA}

\affiliation{$^{2}$ Departamento de F\'\i sica Te\'orica, Universidad 
de Valladolid, Valladolid, SPAIN}

\date{\today}

\begin{abstract}
Several static and dynamic properties of liquid silicon near melting 
have been determined from an orbital free {\em ab-initio} molecular 
dynamics simulation. 
The calculated static structure is in good agreement with the available 
X-ray and neutron diffraction data. The dynamical structure shows 
collective density 
excitations with an associated dispersion relation 
which closely follows recent experimental data. 
It is found that liquid silicon can not sustain the propagation of 
shear waves which can be related to the power spectrum of  
the velocity autocorrelation function. Accurate estimates have also 
been obtained for several transport coefficients. The overall picture is that 
the dynamic properties have many characteristics of the simple liquid metals 
although some conspicuous differences have been found. 
\end{abstract}


\maketitle

\narrowtext

\section{Introduction.}
Molecular dynamics (MD) methods are a 
useful technique to study the properties of liquid 
systems, and the last two decades have witnessed a large increase in
the application of {\it ab-initio} molecular dynamics 
methods (AIMD) based on the density 
functional theory \cite{HK-KS} (DFT) for the treatment of the electron system. 
This theory allows calculation of the ground 
state electronic energy of a collection of atoms, for given nuclear 
positions, and also yields the forces on the 
nuclei via the Hellmann-Feynman theorem. It allows
MD simulations in which the nuclear positions evolve according to 
classical mechanics with the electronic subsystem following adiabatically.  
Most AIMD methods are based on the Kohn-Sham (KS) orbital 
representation of the DFT (KS-AIMD methods) which, at present, 
poses heavy computational demands which limit 
the size of the systems that can be  
studied as well as the simulation times. However, some of these constraints 
are eased in the so-called orbital-free {\it ab-initio} molecular 
dynamics (OF-AIMD) method, 
which by disposing of the electronic orbitals of the KS formulation provides 
a simulation method where the number of variables describing the 
electronic state is greatly reduced, enabling the study of larger 
samples (several hundreds of particles) and for longer 
simulation times (tens of ps).  

This paper reports the results of an {\it ab-initio} molecular 
dynamics simulation on the static and dynamic properties of liquid 
Silicon (l-Si) at thermodynamic conditions near its triple point.  
Si is an interesting material with many peculiar properties 
which, along with its technological importance, has  
stimulated intensive theoretical 
\cite{Stillinger-Weber,Tersoff,Jank-Hafner,Wang,Virkkunen,Stich1,Stich2,Cheli}
and experimental 
\cite{Gabathuler,Waseda1,Waseda2,Takeda,Hosokawa1,Hosokawa2} work. 
Its high-density forms include the crystalline, amorphous and liquid phases 
with the former two being covalently bonded and semiconducting and the latter 
one metallic. At melting Si  
undergoes a semiconductor-metal transition along with a density 
increase of $\approx$ 10\% and significant changes in the local atomic 
structure. This evolves from an open one, with a tetrahedral fourfold 
coordination, to a more compact liquid 
structure with a white-tin-type arrangement preserving the local 
tetrahedrality and with an 
approximate sixfold coordination. \cite{Petkov,Funamori}  

Most theoretical studies on l-Si have focused on its static 
structural properties and have resorted to 
classical MD simulations in which the liquid system 
was characterized by effective 
interatomic potentials constructed either empirically by fitting to 
experimental data \cite{Stillinger-Weber,Tersoff}, or derived 
from some approximate theoretical 
model. \cite{Jank-Hafner,Wang,Virkkunen}. More recently,   
some KS-AIMD calculations \cite{Stich1,Stich2,Cheli} have been 
reported, although the results dealt with electronic and static properties only
because of the computational constraints inherent in this method.

Stich {\it et al} \cite{Stich1} performed the first KS-AIMD calculation for 
l-Si near the triple point using 64 particles, a non-local 
Bachelet-Hamann-Schluter 
type pseudopotential \cite{Bachelet} and the local density approximation (LDA) 
for the electronic exchange and 
correlation energies. Subsequently, a more comprehensive   
study was performed \cite{Stich2} with a larger number of atoms (350 particles) 
and a spin dependent generalized gradient approximation for electron exchange 
and correlation. A KS-AIMD calculation has also been performed by 
Chelikowsky {\it et al} \cite{Cheli} using 64 atoms and a non-local, 
Troullier-Martins type, pseudopotential \cite{TM} These {\it ab-initio} studies 
have provided accurate descriptions of the local liquid structure and valuable 
insights into the the valence electron charge densities, but, by necessity,
they were too short to address dynamical properties 
(besides the self-diffusion coefficient) and 
these properties will now be 
the main focus of the present report which, to our knowledge, is the 
first {\it ab-initio} study on the dynamical properties of l-Si.

The static structure factor, $S(q)$, of l-Si has been 
measured by both neutron scattering (NS) \cite{Gabathuler} and 
X-ray (XR) \cite{Waseda1,Waseda2,Takeda} diffraction. Although all 
experimental  
$S(q)$'s show a distinctive shoulder at the high-$q$ side of the 
main peak, there are some discrepancies in the positions and heights of the 
first peak, its shoulder and the second peak. 
Experimental investigations of the dynamic structure of l-Si have been hampered 
by the large adiabatic sound speed ( $\approx 4000$ m/s),
which prevents the use of the inelastic
thermal neutron scattering (INS) technique for investigating the collective
excitations for small $q$-values. Nevertheless, the dynamical structure of l-Si 
near the triple point has been investigated recently by  
Hosokawa {\it et al} \cite{Hosokawa1,Hosokawa2} using high-resolution 
inelastic X-ray scattering (IXS) which has no such kinematic restrictions.  
The measurements by 
Hosokawa {\it et al} investigated the wavevector regions 
0.02 \AA$^{-1} \leq q \leq $ 1.3 \AA$^{-1}$ [\onlinecite{Hosokawa1}] and  
0.02 \AA$^{-1} \leq q \leq $ 2.96 \AA$^{-1}$ [\onlinecite{Hosokawa2}] 
and several dynamical features already observed in the liquid alkali 
metals were found, 
such as the existence of collective excitations up to $q$-values around 
0.5$\;q_p$ (where $q_p$ is the main peak's position of $S(q)$), along with 
a positive dispersion in the adiabatic sound velocity with respect to the 
hydrodynamic value.

Section \ref{theory} contains a brief description of the 
orbital-free {\em ab-initio} 
molecular dynamics (OF-AIMD) method, and gives some technical details 
particularly on the electronic kinetic energy functional and 
the local pseudopotential used to characterize the ion-electron interaction.  
In section \ref{results} we present and discuss 
the results of the {\em ab-initio} simulations for several static and dynamic 
properties which are compared with the available experimental data. 
Finally, conclusions are drawn and possible avenues for further 
study are suggested.

\section{Theory.}
\label{theory}

A liquid simple metal may be treated as a disordered array of 
$N$ bare ions with valence Z, enclosed in a volume $V$, and 
interacting with $N_{\rm e}=NZ$ 
valence electrons through an electron-ion potential $v(r)$.
The total potential energy of the system can be written, within the 
Born-Oppenheimer approximation, as the sum of the direct ion-ion coulombic 
interaction energy and the ground state energy of the electronic system   
under the external
potential created by the ions, $V_{\rm ext}
(\vec{r},\{\vec{R}_l\}) = \sum_{i=1}^N v(|\vec{r}-\vec{R}_i|)$ ,

\begin{equation}
E(\{\vec{R}_l\}) = \sum_{i<j} \frac{Z^2}{|\vec{R}_i-\vec{R}_j|} +
E_g[\rho_g(\vec{r}),V_{\rm ext}(\vec{r},\{\vec{R}_l\})] \, ,
\end{equation}

\noindent where $\rho_g(\vec{r})$ is the ground state electronic density and 
$\vec{R}_l$ are the ion positions. 
According to DFT, the ground state electronic 
density, $\rho_g(\vec{r})$,  
can be obtained by minimizing the energy functional

\begin{equation}
E[\rho(\vec{r})] = 
T_s[\rho]+ E_H[\rho]+ E_{\rm xc}[\rho]+ E_{\rm ext}[\rho]
\label{etotal}
\end{equation}

\noindent
where the terms represent, respectively, the electron kinetic 
energy, $T_s[\rho]$, 
of a non-interacting system of density $\rho(\vec{r})$, the classical 
electrostatic 
energy of the electrons (Hartree term), 

\begin{equation}
E_H[\rho] = \frac12 \int \int d\vec{r} \, 
d\vec{s} \, \frac{\rho(\vec{r})\rho(\vec{s})}
{|\vec{r}-\vec{s}|} \, ,
\end{equation}

\noindent
the exchange-correlation
energy, $E_{\rm xc}[\rho]$, for which we have used the local density 
approximation 
and finally, the electron-ion interaction energy, $E_{\rm ext}[\rho]$, where 
the electron-ion potential has been characterized by a first principles local 
pseudopotential constructed within DFT \cite{GGLS}. 

\begin{equation}
E_{\rm ext}[\rho] = \int d\vec{r} \, \rho(\vec{r}) V_{\rm ext}(\vec{r}) \, ,
\end{equation}

In the KS-AIMD method, \cite{HK-KS} $T_s[\rho]$ is exactly evaluated by using 
a set of roughly $N_{\rm e}$ single particle orbitals, but at huge 
computational 
expense. In contrast, the OF-AIMD approach uses an approximate but explicit 
density functional for $T_s[\rho]$ so that the system is described in 
terms of the 
single function $\rho(\vec{r})$ replacing the large set of 
orbitals.\cite{HK-KS,
GGLS,Perrot-MaddenLQRT} Proposed functionals incorporate the 
von Weizs\"acker term, 

\begin{equation}
T_W[\rho(\vec{r})] = \frac18 \int d\vec{r} \, 
|\nabla \rho(\vec{r})|^2 /\rho(\vec{r}), 
\end{equation}

\noindent 
plus further terms chosen in order to reproduce correctly some exactly known 
limits. Here, we have used an average density model \cite{GGLS}, where 
$T_s=T_W+T_{\alpha}$, 

\begin{eqnarray}
T_{\alpha} = \frac{3}{10} \int d\vec{r} \, \rho(\vec{r})^{5/3-2\alpha}
\tilde{k}(\vec{r})^2 \\
\tilde{k}(\vec{r}) = (2k_F^0)^3 \int d\vec{s} \, k(\vec{s})
w_{\alpha}(2k_F^0|\vec{r}-\vec{s}|)   \nonumber
\end{eqnarray}

\noindent
$k(\vec{r})=(3\pi^2)^{1/3} \;  \rho(\vec{r})^{\alpha}$, $k_F^0$ is the Fermi 
wavevector for mean electron density $\rho_e = N_e/V$, and $w_{\alpha}(x)$ is a 
weight function chosen so that both the linear response theory and 
Thomas-Fermi limits are correctly recovered. Further details  
are given in Ref. [\onlinecite{GGLS}].

Another key ingredient of the energy functional is the the local ionic 
pseudopotential, $v_{ps}(r)$,  describing the ion-electron interaction.   
It has been constructed from first-principles by fitting, within the same 
$T_s[\rho]$ functional, to the displaced electronic density induced by an ion 
embedded in a metallic medium as obtained in a KS-DFT calculation. Further 
details are given in Ref.  [\onlinecite{GGLS}] and we merely  
note that the 
theoretical framework used in this study has provided an accurate description
of several static and dynamic properties in the  
liquid Li, Mg, Al, Na-Cs and Li-Na systems \cite{GGLS,BGGLS}.

\section{Results}
\label{results}

OF-AIMD simulations have been performed for l-Si  
in a thermodynamic state characterized by  
the temperature T= 1740 K and  ionic  
number density $\rho_i$= 0.05551 \AA$^{-3}$ [\onlinecite{Waseda1}]. 
We have considered 2000 ions in a cubic cell with periodic 
boundary conditions.
Given the ionic positions at time $t$, the 
electron energy functional is minimized with respect to 
$\rho(\vec{r})=\psi(\vec{r})^2$, where $\psi(\vec{r})$ is a single
{\it effective orbital} which is expanded in plane waves 
which are truncated at a cutoff energy, $E_{\rm Cut}=12.75$ Ryd. 
The energy minimization with respect to the 
Fourier coefficients of $\psi(\vec{r})$ is performed 
every ionic time step by using a quenching method which results in the 
ground state electronic density and energy.   
The forces on the ions are obtained from the electronic ground state 
via the Hellman-Feynman theorem, and the ionic positions and velocities are 
updated by solving Newton's equations, with the  
Verlet leapfrog algorithm with a timestep of $3.5\times 10^{-3}$ ps.  
In the simulations equilibration lasted 10 ps. and the calculation of 
properties was made averaging over 50 ps. For comparison we mention that 
the previous KS-AIMD simulations lasted for 1.2 ps. [\onlinecite{Stich1}] 
0.9 ps. [\onlinecite{Stich2}] and 1.0 ps. [\onlinecite{Cheli}], which 
underscores its limitations for dealing with the dynamical properties. 

Several liquid static properties have been evaluated (pair distribution 
function, static structure factor and bond angle distribution) as well 
as various dynamic properties, both single-particle ones (velocity 
autocorrelation function, mean square displacement) and collective ones 
(intermediate scattering functions, dynamic structure factors, 
longitudinal and transverse currents). 
The calculation of the time correlation functions was performed by 
taking time origins every five time steps. Several of these functions  
also depend on wave vector, which for our isotropic system is a dependence
on q$\equiv$$\mid {\bf q} \mid$ only.

\subsection{Static properties.}

\begin{figure}
\begin{center}
\mbox{\psfig{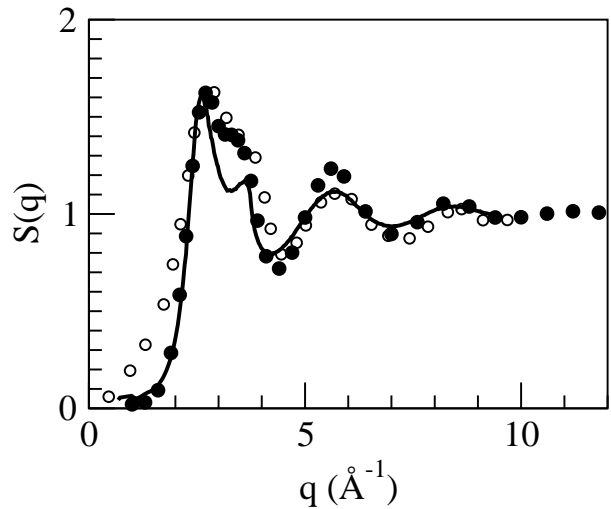}}
\end{center}
\caption{Static structure factor of l-Si at 1740 K. 
Open circles: experimental NS data. \protect\cite{Gabathuler}.   
Full circles: experimental XR 
data. \protect\cite{Waseda1,Waseda2}. 
Continuous line: OF-AIMD simulations.} 
\label{sqfig}
\end{figure}

The pair distribution function, $g(r)$, and the static structure factor $S(q)$
can both be obtained directly from the simulation. The latter 
is plotted in Fig. \ref{sqfig} along with 
the X-ray diffraction data of 
Waseda {\it et al} \cite{Waseda1,Waseda2} and the neutron data of 
Gabathuler and Steeb \cite{Gabathuler}. 

%$$$$$$$$$$$$$$$$$$$$$$$$$$$$$$$$$$$$$$$$$$
%$$$$$$$$$$$$$$$$$$$$$$$$$$$$$$$$$$$$$$$$$$

The OF-AIMD  $S(q)$ has a main peak at $q_p$ $\approx$ 2.59 \AA$^{-1}$ 
with a distinctive shoulder at $\approx$ 3.20 \AA$^{-1}$, 
the second peak is at $\approx$ 5.65 \AA$^{-1}$ and the subsequent 
oscillations are rather weak. As noted earlier there are some  
discrepancies among the measured structures in $q_p$, $S(q_p)$, 
and the positions of the shoulder and the second peak.  
The experimental results for $q_p$ range from 2.49 \AA$^{-1}$ 
[\onlinecite{Takeda}] to 2.78 \AA$^{-1}$ [\onlinecite{Gabathuler}], 
the position of the shoulder from 3.15 \AA$^{-1}$ [\onlinecite{Takeda}] to 
3.40 \AA$^{-1}$ [\onlinecite{Gabathuler}], and the second peak's position from
5.53 \AA$^{-1}$ [\onlinecite{Takeda}] 
to 5.70 \AA$^{-1}$ [\onlinecite{Gabathuler}]. 
In all cases the smaller estimate is from Takeda's XR data \cite{Takeda}
and the larger from the NS data of Gabathuler and Steeb 
\cite{Gabathuler}. It is worth mentioning that the XR  
data by Waseda {\it et al} \cite{Waseda1,Waseda2} gave values  
closer to the NS ones, namely  $q_p$=2.76 \AA$^{-1}$ with 
the shoulder at 3.35 \AA$^{-1}$ 
and the second peak's position at 5.66  \AA$^{-1}$.  
Nevertheless, the values from the OF-AIMD simulation always lie 
within the range covered by the experimental data.

The OF-AIMD $S(q)$ accounts for the position and height of 
the main peak, although the amplitudes of the following peaks and 
troughs are somewhat underestimated. Moreover, the shoulder on the 
high-$q$ side of the main peak, which is the most distinctive feature of 
the measured $S(q)$ for l-Si, is also reproduced by the simulation 
although with a 
smaller height. A similar shortcoming is visible in the $S(q)$ from the 
KS-AIMD simulations of Stich {\it et al} \cite{Stich1,Stich2}, whereas that of 
Chelikowsky {\it et al} \cite{Cheli} gives a shoulder of similar 
height as the main peak.

Extrapolation of $S(q)$ to $q \to 0$ and use of the relation 
$S(0) = \rho_i k_B T \kappa_T$ gives an estimate for the isothermal 
compressibility $\kappa_T$. A least squares fit of $S(q)=s_0+s_2 q^2$ 
to the calculated $S(q)$ for $q$-values  up to 0.8 
\AA$^{-1}$ yields $S(0)$= 0.04 $\pm$ 0.005 and 
$\kappa_{T}$ = 3.0 $\pm$ 0.6
(in 10$^{-11}$ m$^2$ Nw$^{-1}$ units)  for T = 1740 K. 
This latter value stands remarkably close to the experimental 
result \cite{Baidov&Gitis} 
of 2.8 x 10$^{-11}$ m$^2$ Nw$^{-1}$ and therefore 
provides confidence in the small-$q$ behaviour of our calculated 
$S(q)$.

The simulation directly gives the pair distribution function, $g(r)$, whose 
main peak's position, $r_p$, is usually identified with the 
average nearest neighbor distance. The OF-AIMD simulation gives $r_p \approx$ 
2.43 \AA, which is slightly smaller than the experimental data of 
2.45 \AA [\onlinecite{Waseda1,Waseda2}], 
2.47 \AA [\onlinecite{Takeda}] and 
2.50 \AA  [\onlinecite{Gabathuler}]. 
An estimate of the number of nearest neighbors is obtained by integrating the 
radial distribution function (RDF), $4\pi r^2 \rho_i g(r)$, up to a distance 
$r_m$ taken as the position of its 
first minimum \cite{Cusak,McGreevy}; the present OF-AIMD study gives 
$r_m$ $\approx$ 2.95 \AA, yielding to a coordination number
of $\approx$ 6.0 atoms,  which reasonably compares with the 
experimental values at T=1735 K ranging from  $\approx$ 6.4 
atoms \cite{Waseda1} to $\approx$ 5.7 atoms \cite{Takeda}. 

%$$$$$$$$$$$$$$$$$$$$$$$$$$$$$$$$$$$$$$$$$$
\begin{figure}
\begin{center}
%\mbox{\psfig{file=bondangle.eps,angle=0,width=65mm,clip}}
\mbox{\psfig{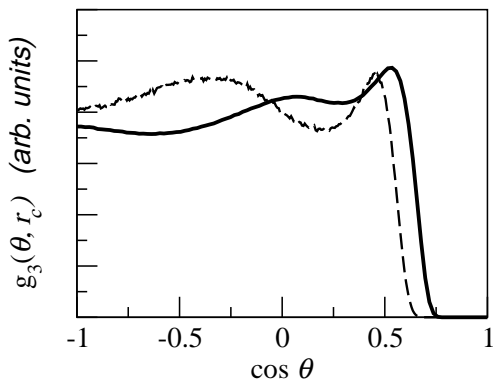}}
\end{center}
\caption{Bond-angle distribution function, $g_3(\theta, r_c)$. 
The cutoff distance $r_c$ is (a) the first minimum of the RDF, 
$r_c$ = 2.95 \AA $\;$ (continuous line) and (b) the covalent cutoff 
$r_c$ = 2.49 \AA $\;$ (dashed line).} 
\label{bondangle}
\end{figure}
%$$$$$$$$$$$$$$$$$$$$$$$$$$$$$$$$$$$$$$$$$$

Signatures of possible remnants of covalent bonding that may remain
in l-Si should be contained in higher order correlation functions. 
A convenient quantity is the bond-angle distribution 
function, $g_3(\theta,r_c)$, where $\theta$ is the angle between two 
vectors joining a reference particle with two neighboring 
particles at a distance less than $r_c$. In liquid simple metals, this 
function peaks at around $\theta$ $\approx$ $60^o$ and $120^o$, which 
are close to those expected for a local icosahedral arrangement 
\cite{Balubook} ($\theta$ $\approx$  $63.5^o$ and $116.5^o$). 
The OF-AIMD results for $g_3(\theta,r_c)$, with $r_c$ taken as the 
first minimum of the RDF, are plotted in Fig. \ref{bondangle} where 
we see two maxima centered around $\theta \approx 60^o$ and $88^o$, which 
practically coincide with the values obtained from the KS-AIMD simulations 
\cite{Stich1,Stich2,Cheli}, namely $\theta \approx 60^o$ and $90^o$. This 
double-peak feature has usually been interpreted as a signature of both 
tetrahedrally bonded and higher coordinated atoms contributing to the 
first coordination shell; 
indeed, according to Stich {\it et al} \cite{Stich1} about 30\% of the 
atoms in the first coordination shell are covalently bonded. 
Also plotted in Fig. \ref{bondangle} is the $g_3(\theta,r_c)$ 
with $r_c$ = 2.49 \AA $\;$ which has been identified \cite{Stich1,Stich2} 
as the valence cutoff distance for covalent bonds. Now, the
OF-AIMD results show a sharp maximum at $\theta$ $\approx$ $65^o$ along with  
broader maximum  centered around the tetrahedral 
angle $\theta$ $\approx$ $109^o$. 
Summing up, the previous results emphasize the 
capability of the OF-AIMD method to account, albeit not so accurately as 
the KS-AIMD methods, for the orientational correlations in those liquid 
systems with some remnants of covalent bonding.

\subsection{Dynamic properties.}

%To find out whether and how 
%the above structural transformation is reflected 
%in the different dynamical magnitudes

\subsubsection{Collective dynamics.}

\begin{figure}
\begin{center}
\mbox{\psfig{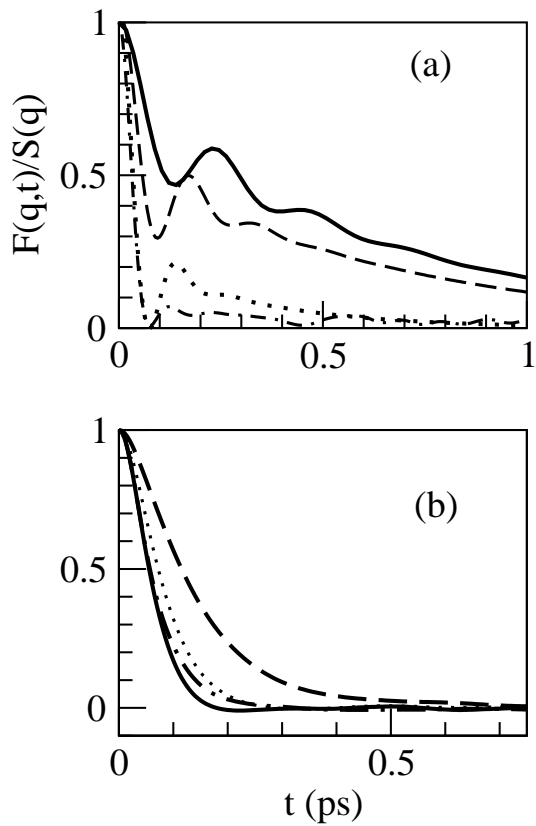}}
\end{center}
\caption{Normalized intermediate scattering functions, $F(q, t)$, at  
several $q$-values (in \AA$^{-1}$ units), for l-Si at T = 1740 K. 
(a) q= 0.46 (full line), 
q=0.72 (dashed line), q=1.07 (dotted line) and  
q=1.46 (dot-dashed line).  (b)  q=2.0 (full line), q=2.57 (dashed line), 
q=3.0 (dotted line) and q=3.62 (dot-dash line).} 
\label{Fqt}
\end{figure}

The intermediate scattering function, $F(q, t)$, defined as
\begin{equation}
F(q, t) = \frac{1}{N} \left \langle \left( \sum_{j=1}^N
e^{-i {\vec q}{\vec R}_j(t + t_0)} \right)
 \; \left( \sum_{l=1}^N e^{i {\vec q}{\vec R}_l(t_0)} \right) \right \rangle
\end{equation}
contains spatial and temporal information on the collective dynamics of 
density fluctuations. Results for $F(q, t)$ as functions of time for
for several $q$-values are shown in Fig. \ref{Fqt}. 
$F(q, t)$ oscillates 
up to $ q \approx (3/5) \; q_p =1.55$\AA$^{-1}$, with the amplitude of 
the oscillations being stronger for the smaller $q$-values. This 
behaviour, which is plotted 
in Fig. \ref{Fqt}(a), is typical of simple liquid metals near melting, 
as found from both computer 
simulations \cite{GGLS,TorBalVer,Shimojo2,Kambayashi} 
and theoretical models. \cite{Litio} 
%**************************************************
%************************************************************
However, at low $q$-values ($q$ $\leq$ 0.5 $q_p$) the $F(q, t)$  
shows a strong diffusive component which plays  a 
dominant role and imposes a slow decay. This is at variance with 
the results for the simple liquid metals near melting, 
\cite{GGLS,TorBalVer,Shimojo2,Kambayashi,Litio} where for a comparable  
$q$-range the diffusive component is already very weak and the 
corresponding $F(q, t)$ shows marked oscillations around zero. 

This different behaviour can be explained in terms of the hydrodynamic 
expression for the $F(q, t)$, which is exact in the $q$ $\to$ 0 
limit \cite{Cusak,Balubook} (hydrodynamic limit)

\begin{eqnarray}
\label{fqt}
& & F(q, t)/S(q) = \left(\frac{\gamma-1}{\gamma}\right) 
\; exp(-D_T(q) t) + \nonumber \\
& &
\frac{1}{\gamma} \; exp(-\Gamma(q) t) \; \left[ \cos{(c_s q t)} + 
b q \sin{(c_s q t)} \right]
\end{eqnarray}

\noindent and its time FT, the dynamic structure factor, $S(q, \omega)$, 

\begin{eqnarray}
\label{sqw}
& & 2 \pi \;  S(q, \omega)/S(q) =  
\left ( \frac{\gamma - 1}{\gamma} \right) 
\frac{2 D_T(q)}{\omega^2 + (D_T(q))^2} + \nonumber \\
& &
\frac{1}{\gamma} 
\left( \frac{\Gamma(q)+bq(\omega+c_sq)}{(\omega+c_sq)^2+ (\Gamma(q))^2} +
\frac{\Gamma(q)+bq(\omega-c_sq)}{(\omega - c_sq)^2+ (\Gamma(q))^2} \right) 
\nonumber \\  
& &
\end{eqnarray}

\noindent 
where $\gamma=C_p/C_v$ is the ratio of specific heats, 
$c_s$ is the adiabatic sound velocity, 
$\Gamma(q)= \Gamma q^2$ with $\Gamma$ the 
sound attenuation constant and 
$D_T(q)= D_T q^2$ where 
$D_T=\kappa_T/(\rho_i C_p)$ is the thermal diffusivity and  
$\kappa_T$ is the thermal conductivity.
Finally, $\Gamma=\frac{1}{2}[a (\gamma-1)/\gamma + \nu_l]$ 
with $a=\kappa_T/(\rho_i C_v)$ and $\nu_l$ is the kinematic longitudinal 
viscosity.

%**************************************************
\begin{figure}
\begin{center}
\mbox{\psfig{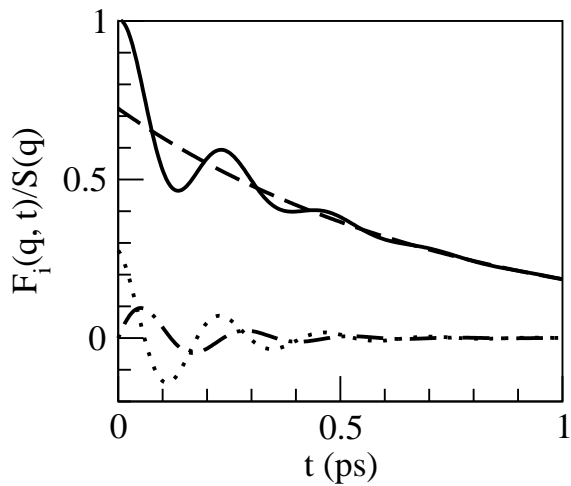}}
\end{center}
\caption{Contributions to the $F(q, t)$ at $q$ =0.47 \AA$^{-1}$ for 
l-Si at T=1740 K. 
Diffusive component (dashed line), cosine component (dotted line) and  
sine component (dash-dotted line).} 
\label{fkti}
\end{figure}
%************************************************************

%**************************************************
\begin{figure}
\begin{center}
\mbox{\psfig{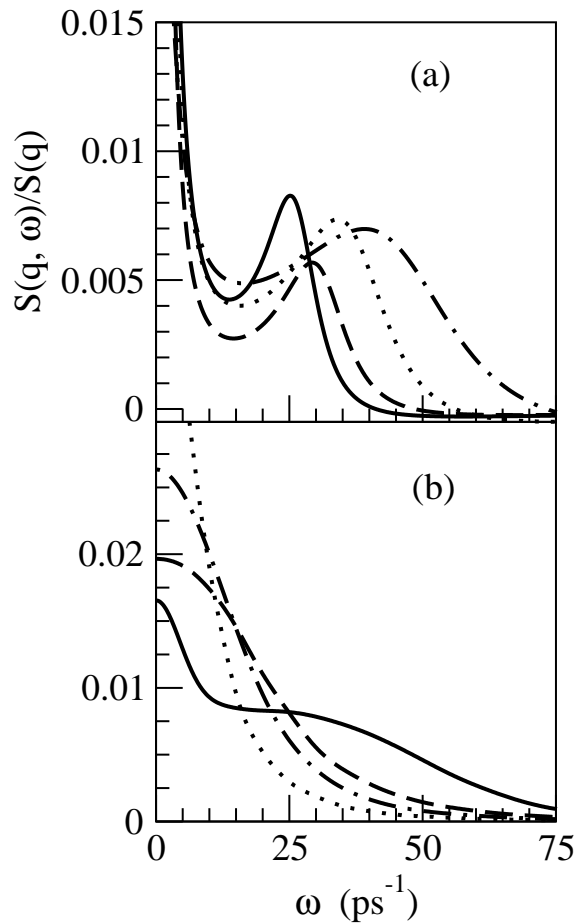}}
\end{center}
\caption{Dynamic structure factor $S(k, \omega)$ 
at several $q$-values (in \AA$^{-1}$ units), for l-Si at T = 1740 K.  
(a) q= 0.46 (full line), 
q=0.58 (dashed line), q=0.72 (dotted line) and  
q=1.07 (dot-dashed line).  
(b)  q=1.46 (full line), q=2.00 (dashed line), 
q=2.57 (dotted line) and q=3.00 (dot-dash line).} 
\label{Sqw}
\end{figure}
%************************************************************

According to Eq. (\ref{sqw}), the hydrodynamic $S(q, \omega)$ 
has two (inelastic) propagating peaks centered at $\omega= \pm c_s q$ 
with half-width at half-maximum (HWHM), $\Gamma(q)$, and 
a diffusive peak at $\omega=0$ and whose width is determined 
by the thermal diffusivity, $D_T$. For a metallic 
system, $D_T$ has electronic and ionic 
contributions with the former dominating; however  
it has been shown \cite{Stroud} that Eqs. (\ref{fqt})-(\ref{sqw}) 
incorporate just that part of $D_T$ due to the ions. Consequently, 
we have fitted the OF-AIMD $F(q,t)$'s at the smallest $q$-values 
($q \leq 0.6$ \AA$^{-1}$), to the analytical expression of Eq. 
(\ref{fqt}). A good fit was achieved in that $q$-range giving an 
approximate value $D_T$ $\sim$ 1.0 x 10$^{-3}$ cm$^2$/sec., which is 
two orders of magnitude smaller than the 
experimental one \cite{Yamamoto}  
$D_T$ = 0.228 $\pm$ 0.004 cm$^2$/sec.  for l-Si at melting, which obviously 
includes both ionic and electronic contributions. We note that an KS-AIMD study 
for liquid Ge near melting \cite{Stroud} gave a comparable estimate of the 
ionic contribution, namely  $D_T$ $\sim$ 1.3 x 10$^{-3}$ cm$^2$/sec., which 
is also two orders of magnitude smaller than its total $D_T$. 
On the other hand, estimates of the ionic contribution to $D_T$ in the liquid 
alkali metals near melting \cite{Cook&Fritsch} range from 20.0 x 10$^{-3}$ 
cm$^2$/sec. (Li) to 3.0 x 10$^{-3}$ cm$^2$/sec. (Cs). Consequently, as $D_T$  
determines the diffusive behaviour of the $F(q,t)$ at small $q$'s 
(see Fig. \ref{fkti}), the larger $D_T$ values of the alkali metals imply 
a much weaker diffusive component which is easily overcome by the 
oscillatory parts of the $F(q, t)$.

$S(q, \omega)$, which is the quantity directly obtained from INS or IXS 
experiments, has been obtained by a time FT of the $F(q, t)$ and 
Fig. \ref{Sqw} shows several $S(q, \omega)$ as functions of $\omega$ for 
different wavevectors up to $\approx$ 1.2$\;q_p$. 
The $S(q, \omega)$ show well defined sidepeaks, indicative of 
collective density excitations, up to $q \approx (3/5)\; q_p$ which 
is also a common feature of the liquid simple metals \cite{GGLS,Balubook}.

%$$$$$$$$$$$$$$$$$$$$$$$$$$$$$$$$$$$$$$$$$$
\begin{figure}
\begin{center}
\mbox{\psfig{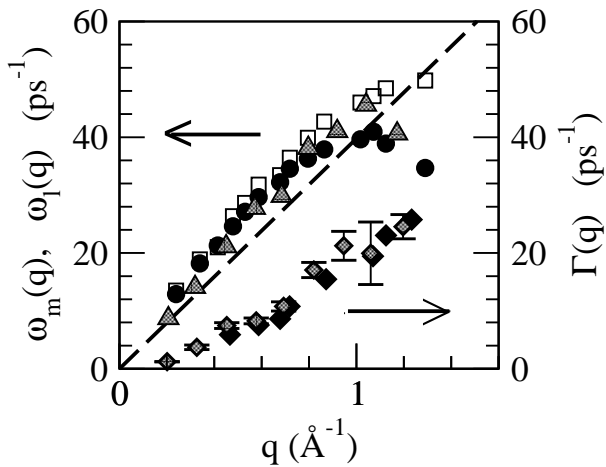}}
\end{center}
\caption{Dispersion relation for l-Si at T =1740 K. Full circles: 
peak positions, $\omega_m(q)$, from the calculated $S(q, \omega)$. 
Triangles: experimental $\omega_m(q)$, from Hosokawa {\it et al} 
(Ref. \onlinecite{Hosokawa1}). 
Open squares: peak positions, $\omega_l(q)$, from the maxima of the  
calculated longitudinal current, $J_l(q, \omega)$. Dashed line: 
Linear dispersion with the hydrodynamic sound velocity, v=3977 m/s. 
Grey (full) diamonds with error lines: experimental (calculated) values for the 
HWHM, $\Gamma(q)$, of the inelastic peaks. }
\label{disperfig}
\end{figure}
%$$$$$$$$$$$$$$$$$$$$$$$$$$$$$$$$$$$$$$$$$$

The dispersion relation of the density fluctuations, $\omega_m(q)$, has been 
obtained from the positions of the sidepeaks of $S(q, \omega)$, and is shown in
Fig. \ref{disperfig}, along with $\omega_l(q)$, which is 
the dispersion relation obtained from the maxima of the longitudinal 
current correlation function, $J_l(q, \omega) = \omega^2 S(q, \omega)$. 
Figure \ref{disperfig} also includes the 
experimental $\omega_m(q)$ data of Hosokawa 
{\it et al} \cite{Hosokawa1} and a line representing the dispersion of the 
hydrodynamic sound whose slope gives the 
experimental \cite{Yoshimoto1} adiabatic 
sound velocity $c_s$ = 3977 m/s at T=1753 K.   
The experimental $\omega_m(q)$ data has a {\it positive dispersion}, 
i.e. an increase of $\omega_m(q)$ with respect to the linear hydrodynamic 
dispersion relation value, with a maximum located 
around 0.8 \AA$^{-1}$ and whose 
magnitude amounts to $\approx$ 15\%. Although there is fair agreement 
between our calculated $\omega_m(q)$ dispersion relation and experiment, the 
uncertainty at the small $q$-values prevents us from unequivocally declaring  
positive dispersion. A similar {\it positive dispersion} has 
been experimentally 
observed in the liquid alkali metals \cite{Scopigno,BurkelSinn,Pilg0}, 
Mg \cite{Kawakita},  Al \cite{Scopigno} and  Hg \cite{Hosokawa3}. 
On the other hand, we recall that at the hydrodynamic region, the slope 
of the dispersion relation curve gives a $q$-dependent adiabatic sound 
velocity, $c_s(q)$,  which in the limit $q \to 0$ reduces to the 
bulk adiabatic sound velocity.

Another quantity characterizing the collective density excitations is 
the HWHM, $\Gamma(q)$, proportional to the inverse lifetime of the  
excitations. Figure \ref{disperfig} shows the experimental data for   
$\Gamma(q)$, which at low $q$ values is $\propto q^2$ (hydrodynamic limit)
but departs from it as $q$ increases \cite{Hosokawa1,Hosokawa2}.     
The figure also includes the $\Gamma(q)$ obtained from the simulation, 
which agrees fairly well with the experiment. In passing we note that to 
calculate $\Gamma(q)$, the $F(q, t)$ obtained from the simulation is fit 
to an eight-parameter analytical expression that interpolates among 
the ideal gas, viscoelastic and hydrodynamic models \cite{Ebbsjo}.   
This method allows the different contributions to $S(q, \omega)$ to be
disentangled and gives an estimate of $\Gamma(q)$ which in the present 
calculations is within an error of $\approx$ 15 \%.

\medskip

\begin{figure}
\begin{center}
%\mbox{\psfig{file=Sqw.eps,angle=0,width=75mm,clip}}
\mbox{\psfig{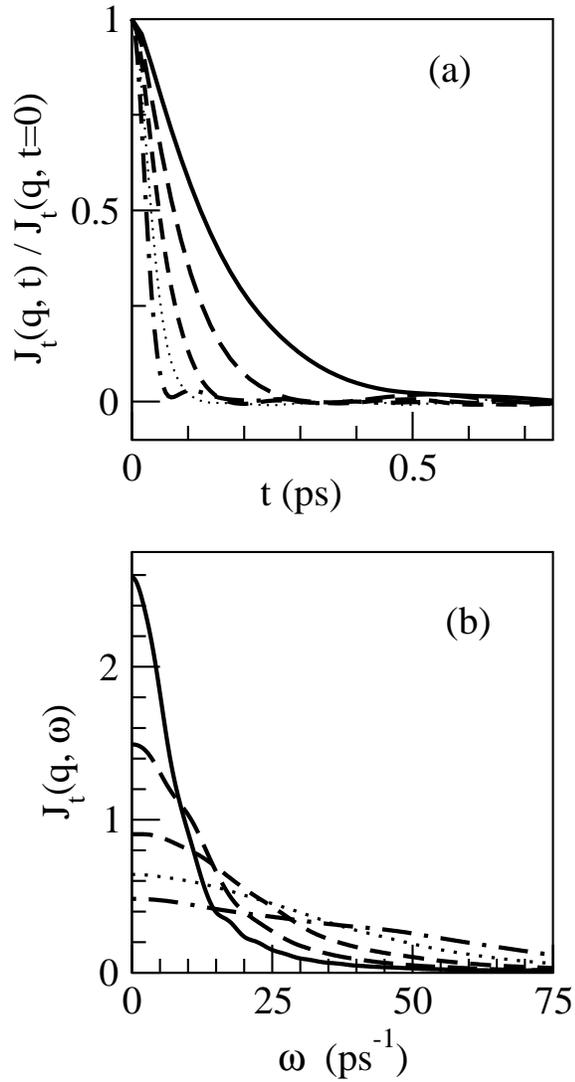}}
\end{center}
\caption{(a) Transverse current correlation function $J_t$(q,t) 
at several $q$-values (in \AA$^{-1}$ units), for l-Si at T = 1740 K.  
q= 0.46 (full line), q=0.72 (long dashed line),  
q=1.07 (short-dashed line),  q=1.46 (dotted line) and 
q=2.57 (dash-dotted line). (b) Same for 
$J_t$(q,$\omega$). } 
\label{Ctqtw}
\end{figure}

The transverse current correlation function, $J_t(q,t)$, is not associated with 
any measurable quantity and can only be determined from computer 
simulations. It 
provides information on the shear modes and its shape evolves from a gaussian, 
in both $q$ and $t$, at the free particle ($q \to \infty$) limit, to a gaussian 
in $q$ and exponential in $t$ at the hydrodynamic limit ($q \to 0$), i.e. 

\begin{equation}
J_t(q \to 0, t) = \frac{1}{\beta m} e^{-q^2 \eta \mid t \mid /m \rho_i} \; ,
\label{Jtqthyd}
\end{equation}

\noindent 
where $\eta$ is the shear viscosity coefficient, $\beta=(k_B T)^{-1}$ and  
$m$ is the atomic mass. In both the above limits $J_t(q, t)$ is positive, but
at intermediate $q$-values there is usually a more complicated behavior with 
well-defined oscillations within a 
limited $q$-range \cite{GGLS,Balubook,Boon&Yip}. 
$J_t(q, t)$ as functions of time for l-Si have been obtained 
from the simulation 
and are shown in Figure \ref{Ctqtw} for several $q$-values up to $q_p$. The
most striking feature is the extreme weakness of the oscillations around zero. 
This is in remarkable contrast to the enormous amount of MD results for 
different liquid systems such as hard sphere 
systems \cite{Boon&Yip}, Lennard-Jones 
liquids \cite{Balubook,Boon&Yip} and simple liquid metals 
\cite{GGLS,Balubook,Kambayashi} near melting, where 
the $J_t(q, t)$ exhibit strong 
oscillations and the associated spectra, $J_t(q, \omega)$, have 
an inelastic peak 
which appears at low $q$-values and lasts for a finite $q$-range. 
Again, l-Si near 
melting shows a different behaviour  with the corresponding $J_t(q, \omega)$, 
which is also plotted in Figure \ref{Ctqtw}, exhibiting a monotonic 
decreasing behaviour 
at all $q$'s. Taking into account that the inelastic peak in 
the $J_t(q, \omega)$, 
is associated with propagating shear waves, the results above point to their 
absence in l-Si near melting. 

An estimate of the shear viscosity coefficient, $\eta$, has been made from 
$J_t(q, t)$ as follows. \cite{GGLS,Palmer,BaBroJedVa} The memory function 
representation of $J_t(q,  t)$:  

\begin{equation}
\tilde{J_t}(q, z)= \frac{1}{\beta m} 
\left [ z + \frac{q^2}{\rho_i m} \;  \tilde{\eta}(q, z)\right ]^{-1} \; ,
\end{equation}

\noindent where the tilde denotes the Laplace transform, introduces a  
generalized shear viscosity coefficient, $\tilde{\eta}(q, z)$. The area under 
the normalized $J_t(q, t)$, gives  $\beta m \; \tilde{J_t}(q, z=0)$ from 
which values 
for $\tilde{\eta}(q, z=0)$ can be obtained and these may be 
extrapolated to $q=0$ 
to give the usual shear viscosity coefficient, $\eta$. The OF-AIMD simulations 
give $\eta$=0.75 $\pm$ 0.1 GPa $\cdot$ ps, which fairly compares with 
the available experimental 
results \cite{Sasaki-Kimura} $\eta_{exp}$=0.58-0.78  GPa $\cdot$ ps.

%**************************************************
%************************************************************

\subsubsection{Single-particle dynamics.}

\begin{figure}
\begin{center}
\mbox{\psfig{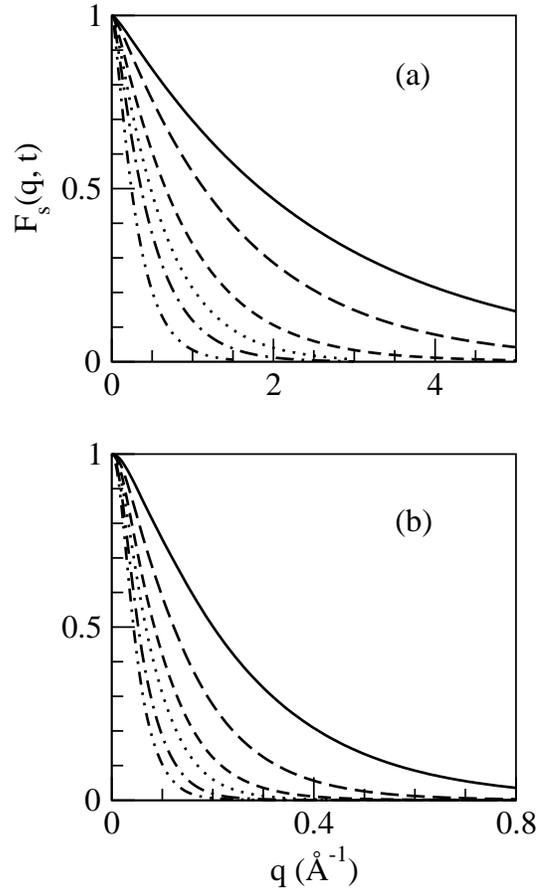}}
\end{center}
\caption{Self intermediate scattering functions, $F_s(q, t)$, at  
several $q$-values (in \AA$^{-1}$ units), for l-Si at 
1740 K. (a) q= 0.42 (full line), 
q=0.54 (dashed line), q=0.72 (short dashed line), 
q=0.86 (dotted line), q=1.02 (dot-dash line) and , 
q=1.29 (doble dot-dashed line). 
(b) q= 1.46 (full line), 
q=2.0 (dashed line), q=2.57 (short dashed line), 
q=3.0 (dotted line), q=3.6 (dot-dash line) and , 
q=4.2 (doble dot-dashed line).} 
\label{fsqtfig}
\end{figure}

The most comprehensive information on the single-particle properties is 
provided by the self-intermediate scattering 
function, $F_s(q, t)$, which probes 
the single-particle dynamics over different length scales, ranging 
from the hydrodynamic to the free-particle 
limit. In the present simulations, this quantity has been 
obtained using 

\begin{equation}
F_s(q, t) = \frac{1}{N} \langle \sum_{j=1}^N  
e^{-i {\vec q}{\vec R}_j(t + t_0)} 
  e^{i {\vec q}{\vec R}_j(t_0)} \rangle
\end{equation}

\noindent 
and Fig. \ref{fsqtfig} depicts the results obtained for several $q$-values.  
It shows the usual monotonic decay with time; but comparison with 
the liquid simple metals (i.e. alkali, alkali-earths, 
Al) \cite{GGLS,Balubook,Litio} 
near their triple point shows that at similar $q/q_p$ values, 
the $F_s(q, t)$ for
l-Si decays much faster. As explained later, this fast decay is related to the 
significantly greater self-diffusion coefficient in l-Si.   

%****************************************
%*******************************************************

Closely related to $F_s(q, t)$ is the velocity autocorrelation function (VACF)
of a tagged ion in the fluid, $Z(t)$, which can be obtained as the $q \to 0$ 
limit of the first-order memory function of the $F_s(q, t)$. However, in the 
present simulations it is more easily obtained from its definition

\begin{equation}
Z (t) = \langle \vec{v}_1(t) \vec{v}_1(0) \rangle
/ \langle v_1^2 \rangle  
\end{equation}

\begin{figure}
\begin{center}
\mbox{\psfig{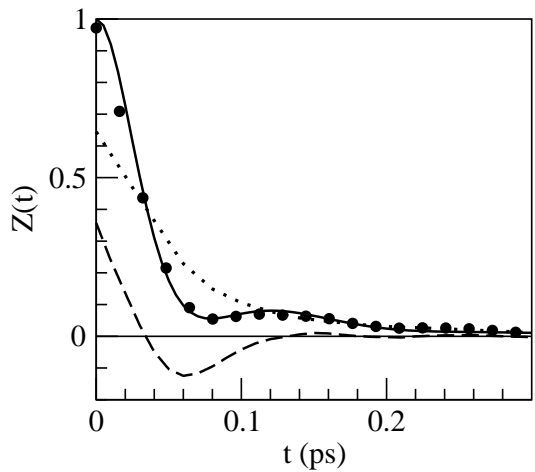}}
\end{center}
\caption{ Normalized OF-AIMD $Z(t)$ for 
l-Si at 1740 K (full line). The dashed and dotted lines are the 
longitudinal, $Z_l(t)$, and transverse, $Z_t(t)$, components respectively, 
as defined in Eq. (\ref{Ztmc}), and the full circles represent their sum. }
\label{vacf-t}
\end{figure}

\begin{figure}
\begin{center}
\mbox{\psfig{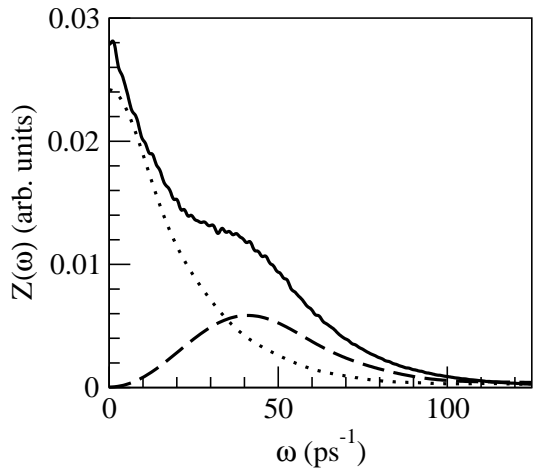}}
\end{center}
\caption{ Same as in the previous figure but for the 
corresponding power spectrum $Z(\omega)$. } 
\label{vacf-w}
\end{figure}

Figures \ref{vacf-t}-\ref{vacf-w} show, respectively, 
the $Z(t)$ and its power 
spectrum $Z(\omega)$ obtained by taking the Fourier 
transform (FT) of the $Z(t)$. 
Besides the expected diffusive modes at low frequencies, the $Z(\omega)$ shows 
a shoulder at $\omega$ $\approx$ 35 ps$^{-1}$, which 
is close to the optical vibrational frequency of covalent Si, and  
it has been related to the existence of vibrational modes echoing 
the remnant covalent bonds in the liquid 
\cite{Stich1}.

%$$$$$$$$$$$$$$$$$$$$$$$$$$$$$$$$$$$$$$$$$$
%$$$$$$$$$$$$$$$$$$$$$$$$$$$$$$$$$$$$$$$$$$
%$$$$$$$$$$$$$$$$$$$$$$$$$$$$$$$$$$$$$$$$$$
%$$$$$$$$$$$$$$$$$$$$$$$$$$$$$$$$$$$$$$$$$$

The $Z(t)$ shown in Fig.  \ref{vacf-t} lacks the usual backscattering 
behaviour observed in the liquid simple metals near melting which is
usually ascribed to the so-called "cage effect" due to the shell of 
nearest neighbors. The relatively open structure of l-Si with $\approx 6$ 
nearest neighbors gives rise to a negligible "cage effect" and $Z(t)$ goes 
monotonically to zero. Indeed, its shape is remarkably similar to that 
from the KS-AIMD calculations of Stich {\it et al} \cite{Stich1,Stich2}. 
The self-diffusion coefficient, $D$, is readily obtained from either 
the time integral of $Z(t)$ or from the slope of the mean square 
displacement $\delta R^2(t) \equiv \langle | \vec{R}_1(t) - \vec{R}_1(0) 
|^2 \rangle$ of a tagged ion in the fluid, as follows   

\begin{equation}
D= \frac{1}{\beta m} \int_0^{\infty} Z(t) dt\; ;\hspace{0.91 cm}
D= \lim_{t \to \infty} \delta R^2(t)/6t 
\end{equation}

\noindent
Both routes lead to practically the same value 
for $D_{\rm OF-AIMD}= 2.28 \;$ \AA$^2$/ps.  
We know of no experimental results for the diffusion coefficients 
of l-Si at any 
thermodynamic state, but note that the KS-AIMD calculations 
of Stich {\it et al} \cite{Stich1,Stich2} gave  
$D_{\rm KS-AIMD}=2.02 \;$ \AA$^2$/ps. which slightly 
increased to 2.4  \AA$^2$/ps. when the number 
of particles was increased to 350 particles. They found 
\cite{Stich2} that a spin 
dependent treatment of electron exchange and correlation within the 
GGA led to a further increase to $D_{\rm KS-AIMD}= 3.1 \;$ \AA$^2$/ps., 
which was explained by a weakening of the 
interatomic bonds in comparison with the LDA treatment. 
The KS-AIMD study of Chelikowsky {\it et al} \cite{Cheli} yielded 
$D_{\rm KS-AIMD}= $1.90 $\;$ \AA$^2$/ps.
The present OF-AIMD result remains within the set of values predicted 
by the KS-AIMD method. Other calculations based on tight binding or 
interatomic pair potentials \cite{Stillinger-Weber,Wang,Virkkunen} tend to 
produce significatively smaller values for the self-diffusion coefficient  
along with a VACF which takes negative values. 
We note that {\it ab-initio} estimates of the self-diffusion coefficient of 
l-Si are nearly one order of magnitude greater than corresponding ones
for the liquid simple metals (alkalis, alkali-earths, Al) near 
melting \cite{GGLS,Iida,Alemany}. This larger $D$ explains why 
$F_s(q,t)$ decays, for any $q$-value, much faster than in the simple 
liquid metals. The accurate gaussian 
approximation \cite{Balubook,Boon&Yip} gives   
$F_s(q,t)= exp [ - q^2 \; \delta R^2(t)  /6 ]$, and a greater $D$ implies a 
larger $\delta R^2(t)$ and a faster decay for the $F_s(q,t)$.

With knowledge of the self-diffusion coefficient $D$ and regarding the motion 
of an atom as the Brownian motion of a macroscopic particle of diameter $d$, 
the Stokes-Einstein (SE) relation, $\eta$ $D$ = $k_B T$/2$\pi$$d$ can be
used to obtain an estimate for the viscosity $\eta$. Identifying $d$ with the 
position of the main peak of $g(r)$ as is commonly done 
gives $d=2.40$ \; \AA $\;$ and 
taking $D_{\rm OF-AIMD}= 2.28$  \AA$^2$/ps leads to $\eta$=0.69 GPa $\cdot$ ps 
which is consistent with the previous OF-AIMD estimate obtained 
from the transverse current correlation function. 

To achieve a deeper physical insight into 
the features of $Z(t)$, we resort to the mode-coupling (MC) approximation 
of Gaskell and Miller \cite{Gaskell&Miller} which provides information 
about the relative importance of the coupling of the single-particle 
motion to the collective longitudinal and transverse currents. It has 
already been used used to interpret MD data 
obtained for the VACF in Lennard-Jones fluids, Yukawa fluids, alkali metals 
and even some molecular 
liquids \cite{Gaskell&Miller,Barrat,Baluwater}. Within this approach

\begin{eqnarray}
\label{Ztmc}
Z (t) 
& & \approx \frac{1}{24 \pi^3} \int d{\bf q} \; f(q) 
\left[ J_l(q, t) + 2 J_t(q, t) \right] F_s(q, t) \nonumber \\ 
& & 
\equiv Z_l(t) \;  + \;  Z_t(t) \nonumber \\
\end{eqnarray}

\noindent where $J_l(q, t)$ and $J_t(q, t)$ are 
the normalized longitudinal and transverse current correlation 
functions and  

\begin{equation}
f(q)= \frac{3}{\rho_i} \frac{j_1(aq)}{aq}
\end{equation}

\noindent 
where $j_1(x)$ is the spherical Bessel function of order one   
and $a$ is an ionic radius given by $(4/3)\pi a^3 = 1/\rho_i$. 
Using the OF-AIMD results for $J_l(q, t)$, $J_t(q, t)$ and $F_s(q, t)$  
in Eq. (\ref{Ztmc}) allows the contributions of the longitudinal 
and transverse currents to be separated as  
$Z_l(t)$ and $Z_t(t)$ respectively. Firstly we note that, as shown in 
Fig. \ref{vacf-t}, the $Z(t)$ obtained by applying 
Eq. (\ref{Ztmc}) is in excellent agreement with the $Z(t)$  
given by the OF-AIMD simulation; therefore it may be reliably used to
get deeper insight into the behavior of the $Z(t)$ by analyzing the 
separate contributions $Z_l(t)$ and $Z_t(t)$ to the integral (\ref{Ztmc}).  

These contributions to $Z(t)$ are plotted in Fig. \ref{vacf-t} which 
shows that $Z_l(t)$ has some oscillatory behaviour, but $Z_t(t)$ remains 
positive for all times as a consequence of the mostly positive values taken 
by the $J_t(q,t)$'s (see Fig. \ref{Ctqtw}). The positive bump of $Z(t)$ 
arises from the fast decay of $Z_l(t)$ which after a negative minimum at 
a rather short time ($\approx$ 0.06 ps) quickly goes to zero. This behaviour, 
along with the comparatively slower decay of $Z_t(t)$ leads to the bump at 
$\approx$ 0.12 ps. At longer times, the dynamics of $Z(t)$ is completely 
determined by $Z_t(t)$. This shape has strong similarities with the MD 
simulation results for liquid water at room conditions \cite{Baluwater}, 
although the $Z(t)$ of l-Si always remains positive whereas in 
liquid water it takes negative values after the bump. However, in both systems 
the $Z(t)$ is strongly influenced by $Z_t(t)$. 
In contrast, the simple liquid metals show a very different pattern 
\cite{Gaskell&Miller} with both $Z_l(t)$ and $Z_t(t)$ oscillating about
zero. The $Z(t)$ also shows oscillations which, beyond the first minimum, 
are primarily determined by the $Z_l(t)$ which also dominates the large $t$ 
behaviour of $Z(t)$.

The longitudinal and transverse components of the power spectrum are 
shown in Fig. \ref{vacf-w}. The small-$\omega$ 
behaviour is completely dominated by the $Z_t(\omega)$, whereas the shoulder 
in the $Z(\omega)$ is induced by the 
maximum in $Z_l(\omega)$ at $\omega$ $\approx$ 40 ps$^{-1}$. 
%at $\omega$ $\approx$ 35 ps$^{-1}$ coincides with the position of the 
%maximum in $Z_l(\omega)$. 
Note that the absence of a peak in $Z_t(\omega)$ 
can be traced back to the lack of inelastic peaks in $J_t(q, \omega)$ 
(see Fig. \ref{Ctqtw}), which is in turn an expression of the failure 
of l-Si to sustain shear waves. 

Again, there are substantial differences between the dynamical results for
l-Si and those for liquid simple metals for which both $Z_t(\omega)$ and 
$Z_l(\omega)$ have clear maxima at $\omega_t^m$ and $\omega_l^m$ with 
$\omega_t^m$ $\le$ $\omega_E$ $\le$ $\omega_l^m$, where $\omega_E$ is the 
so-called "Einstein frequency" of the metal.\cite{Balubook}  
Consequently, the $Z(\omega)$ for liquid metals usually has a maximum 
at $\approx$ $\omega_t^m$ and a shoulder at $\approx$ $\omega_l^m$ which, 
incidentally, is also the pattern exhibited by the MD results for liquid water 
\cite{Baluwater}. In their study of the $Z(t)$ for simple liquids 
\cite{Gaskell&Miller}, Gaskell and Miller, suggested that the appearance of a 
peak in $Z(\omega)$ at a frequency well below $\omega_E$ was indirect evidence 
that the liquid system can sustain the propagation of shear waves. The present 
results which show that l-Si lacks both the frequency peak and shear waves, are
consistent with their suggestion. Finally, we note that the diffusion 
coefficient, 
which is proportional to $Z(\omega=0)$, is completely determined by the 
transverse component which contributes  $\approx$ 99 \% of the total.

%***********************************************
\begin{figure}
\begin{center}
\mbox{\psfig{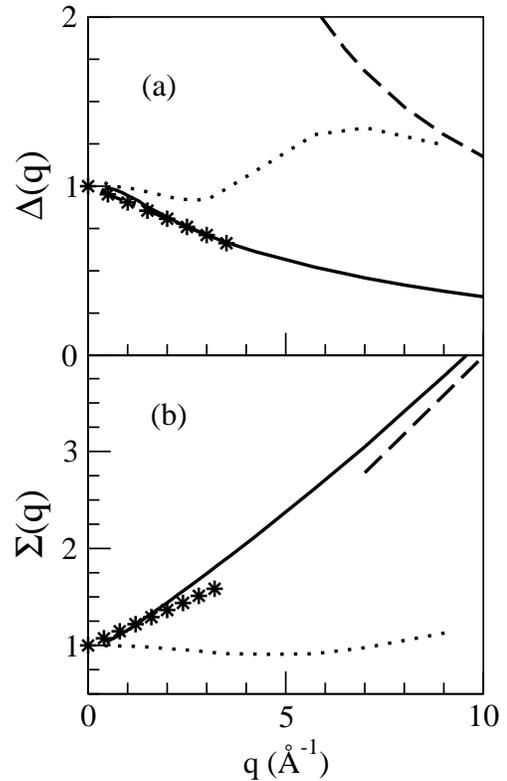}}
\end{center}
\caption{(a) Normalized HWHM of $S_s(q, \omega)$, relative to 
its value at the hydrodynamic limit, for l-Si at T = 1740 K  
(continuous line) and liquid Al near melting (dotted line). 
The asteriks are the predictions of the mode-coupling theory and the 
dashed line gives the free-particle limit.  (b) Same as above, but for 
the normalized peak value $S_s(q, \omega=0)$, relative to 
its value at the hydrodynamic limit. }
\label{DelSig}
\end{figure}
%%**************************************************************

The  time FT of $F_s(q, t)$ gives its frequency spectrum, $S_s(q, \omega)$, 
known as the self-dynamic structure factor, which has experimental 
importance due 
to its connection with the incoherent part of the INS cross-section.  
The $S_s(q, \omega)$ exhibits, for all 
$q$-values, a  monotonic decay with frequency from a peak 
value at $\omega = 0$.  
Usually, $S_s(q, \omega)$ is characterized by the peak 
value, $S_s(q, \omega = 0)$, 
and the HWHM, $\omega_{1/2}(q)$, which are frequently reported normalized with 
respect to the values of the hydrodynamic  ($q$ $\to$ 0 ) limit, by introducing 
the dimensionless quantities $\Sigma(q)=\pi  q^2 D S_s(q, \omega =0)$ and 
$\Delta(q) = \omega_{1/2}(q)/ q^2 D$. 
The OF-AIMD results for $\Delta(q)$ and $\Sigma(q)$ are shown in 
Fig. \ref{DelSig} along with OF-AIMD results for liquid Al for comparison. 
Both magnitudes exhibit a behaviour completely different from that of 
liquid Al which, in turn, stands as typical
of the simple liquids near melting. Indeed, it has been found 
\cite{GGLS,Balubook,TorBalVer} that liquid simple metals near 
their triple point have an oscillating $\Delta(q)$, stemming from the 
"cage effect", whereas for a dense gas it decreases monotonically from 
unity at $q$=0 to a $1/q$ behaviour 
at large $q$. The results for l-Si for both $\Delta(q)$ and $\Sigma(q)$ are 
more similar to those for a dense gas, which must be a consequence of the 
open structure. 
An indirect check on the reliability of these results may be provided  
by the MC theory \cite{Sjogren,Sjogren2} which, has already shown 
its capability to account for the experimental 
$\Delta(q)$ and $\Sigma(q)$ in liquid simple
metals \cite{Moont,Cabrillo} at $q$ $\leq$ $q_p$. According to 
the MC theory 

\begin{eqnarray}
\label{modecp}
\Delta(q) = 1 + H(\delta) q/q^* \\ 
\Sigma(q)= 1 + G(\delta^{-1})q/q^*     \nonumber
\end{eqnarray}

\noindent for small $q$-values, where $q^* = 16 \pi m \rho_i \beta D^2$, 
$\delta= D/(D+\eta/m \rho_i)$ and the functions $H(\delta)$ and 
$G(\delta^{-1})$ are given in Ref. \onlinecite{Moont}. 
In both equations the first term is the hydrodynamic result and  
the second one accounts for the coupling of mass diffusion with 
the collective modes. The OF-AIMD results for $D$ and $\eta$ have
been used to determine $\Delta(q)$ and $\Sigma(q)$ given by Eq.  
(\ref{modecp}) and the results are also shown in Fig. \ref{DelSig}. 
For $q$ $\leq$ $q_p$ the OF-AIMD and MC theory results are in good 
agreement, similar to that found for simple metals. This agreement  
points to the ability of the MC theory to describe the single particle 
dynamics, and presumably the collective dynamics too, in l-Si.

\section{Conclusions.}

Several static and dynamic properties of l-Si at a thermodynamic state close 
to the triple point have been evaluated. The simulations have 
been performed using the orbital free {\em ab-initio} molecular dynamics 
method combined with a first-principles local pseudopotential constructed 
within the same framework.

The results obtained for the static structure are comparable to those 
obtained by earlier KS-AIMD calculations \cite{Stich1,Stich2,Cheli}, and 
are in reasonable agreement with the available experimental data. 
We also stress the good description provided by the OF-AIMD method for the 
orientational correlations, as described by the 
bond-angle distribution function, $g_3(\theta,r_c)$. 
The obtained structural magnitudes are in vein with other previous 
{\it ab-initio} studies 
which suggested for l-Si a local structure consisting in a mixture of 
both tetrahedrally bonded atoms and a higher coordinated 
structure (probably a distorted metallic white-tin structure).  
This latter assertion is based on 
the idea \cite{Petkov} that for non-close packed systems, the 
structure of the molten state resembles that of the high pressure 
solid state. If we now recall that in crystalline Si the   
semiconducting diamond structure contracts with pressure and 
transforms at 12 GPa to the metallic white-tin structure, this latter one 
looks as the most likely to be appear in l-Si. 

Although the {\it ab-initio} methods based on the Kohn-Sham approach provide 
a deeper insight into those physical magnitudes related to bonding 
and %detailed 
electronic properties \cite{Stich1,Stich2}, their heavy computational demands 
have precluded their application to the study of the dynamic properties of 
liquids. 
Indeed, the calculation of the dynamical properties in l-Si  
has been the main aim of the present calculations which, incidentally 
were spurred by recent IXS data \cite{Hosokawa1,Hosokawa2}. Moreover, we 
stress that this is the first {\it ab-initio} study on the 
dynamical properties of l-Si. 

The intermediate scattering functions, $F(q, t)$, have at low-$q$ values, 
a strong diffusive component which is comparable to what has been found in 
liquid Ge, but is at variance with the liquid simple metals where the diffusive 
component is much weaker. The dynamic structure factors, $S(q, \omega)$, show 
collective density excitations over a similar range of wavelenths as those in 
liquid simple metals. Moreover, the dispersion relation of the excitations  
closely follows the existing experimental data \cite{Hosokawa1,Hosokawa2}. 

The transverse current correlation functions, $J_t(q, t)$, show extremely 
weak oscillations around zero and take positive values for most of the time. 
This shape leads to spectra, $J_t(q, \omega)$, with no inelastic peaks which 
in turn reflect the absence of shear waves.  
This conclusion is reinforced by the analysis of the spectra of the 
VACF and  appears as an effect of the low coordination number with its 
attendant negligible "cage effect". This is also reflected in the single 
particle dynamics, as embodied in the VACF and the self-intermediate 
scattering functions, leading to a behaviour different from that typical 
of the liquid simple metals near melting. 
Nevertheless, the MC theory appears capable of accounting for the 
single-particle dynamics in l-Si. 
Calculated self-diffusion and shear viscosity transport coefficients were in 
fair agreement with experiment and/or other {\it ab-initio} calculations.  

Up to now, the OF-AIMD method has shown its ability to provide an 
accurate description of the bulk static and dynamic properties in 
simple liquid systems \cite{GGLS,BGGLS}. However, 
the present results for the static and dynamic 
properties of l-Si illustrate the potential of the OF-AIMD method for 
treating liquid systems with remnants of covalent bonding 
and therefore open new venues concerning its range of applicability.

\section*{Acknowledgements}

This work has been supported by the DGICYT of Spain (MAT2002-04393-C0201) and 
the NSERC of Canada. DJG acknowledges additional financial support from  
the Physics Dept. at Queen's University.

\end{document}